\documentclass[11pt]{article}
\usepackage{graphicx}
\usepackage{amssymb,amsmath,amsthm,enumerate,graphics,psfrag}

\newcommand{\comment}[1]{}
\newcommand{\N}{\mathbb N}

\newtheorem{theorem}{Theorem}[section]
\newtheorem{problem}[theorem]{Problem}
\newtheorem{conjecture}[theorem]{Conjecture}
\newtheorem{proposition}[theorem]{Proposition}
\newtheorem{lemma}[theorem]{Lemma}
\newtheorem{corollary}[theorem]{Corollary}

\newcommand{\maj}{\mathit{maj}}

\newcommand{\inst}{\mathit{inst}}
\newcommand{\winst}{\mathit{winst}}
\newcommand{\calH}{\mathcal{H}}
\newcommand{\calQ}{\mathcal{Q}}

\begin{document}

\title{Geodesic stability  for \\ memoryless binary long-lived consensus}

\author{Cristina G. Fernandes\thanks{ Instituto de Matem\'atica e Estat\'{\i}stica, Universidade de S\~ao Paulo, Brazil, \texttt{cris@ime.usp.br}. Partial support by CNPq 309657/2009-1 and 475064/2010-0.}
\ and Maya Stein\thanks{Centro de Modelamiento Matem\'atico, Universidad de Chile, Santiago, Chile, 
\texttt{mstein@dim.uchile.cl}. Support by Fondecyt 11090141 and Fapesp~05/54051-9.}}
 


\maketitle


\begin{abstract}
  The determination of the stability of the long-lived consensus problem is a
  fundamental open problem in distributed systems. We concentrate on the
  memoryless binary case with geodesic paths. We offer a conjecture on the
  stability in this case, exhibit two classes of colourings which attain this
  conjectured bound, and improve the known lower bounds for all colourings. We
  also introduce a related parameter, which measures the stability only for
  certain geodesics, and for which we also prove lower bounds.
\end{abstract}

\section{Introduction}

The {\em consensus problem} in distributed systems consists of the following:
given a set of values, each coming from a processor or sensor, decide on a
representative value, meaning the consensus of the given values.  The
\emph{long-lived} consensus problem consists of repeatedly solving related
instances of the consensus problem. In~\cite{DolevR03}, Dolev and Rajsbaum
introduce the concept of \emph{stability} of long-lived consensus, where one
wishes the representative values, produced by an algorithm for a sequence of
input instances, to change as few times as possible (there might be some cost
associated to a change). So the question is how to choose the outputs in a way
that they are stable in time. In the case with memory, the algorithm may use
the value produced for the previous instances in the sequence to decide on the
value of the current instance. That is not allowed in the so called
\emph{memoryless} case. See also~\cite{EATCS06}.


We will consider binary-valued consensus, with the input sequences being a geodesic path. 
The case with memory is completely solved in~\cite{DolevR03} and also, for the memoryless 
case, some bounds for the minimum number of changes are shown, which we will improve here.
Davidovitch, Dolev, and Rajsbaum~\cite{DavidovitchDR07} consider multi-valued consensus. 
Becker et al.~\cite{BeckerRRR10} study average instability for 
binary consensus using random walks instead of geodesic paths.

\smallskip

We need a few definitions in order to properly state the problem.  
The {\em $n$-hypercube} is $\mathcal H_n:=\{0,1\}^n$. Write $0^n$ for
$(0,0,\ldots,0)$, and similar. The {\em ball $B_t(0^n)$ of radius $t$} around
$0^n$ consists of all elements of~$\mathcal H_n$ with at most~$t$ entries
identical to~$1$. In the same way, we define~$B_t(1^n)$.

A {\em colouring} of $\mathcal H_n$ is a function $f:\mathcal H_n\to \{0,1\}$.
We say that a colouring~$f$ {\em respects} $B_t(0^n)$ and $B_t(1^n)$ if
$f(x)=0$ for each $x$ in $B_t(0^n)$ and $f(x)=1$ for each $x$ in $B_t(1^n)$.
Observe that if $n<2t+1$, the two balls $B_t(0^n)$ and $B_t(1^n)$ intersect, and no colouring can respect $B_t(0^n)$ and $B_t(1^n)$.
As we are not interested in this case, we say $t$ is \emph{valid} (for $n$) if $n \geq 2t+1$.

A {\em geodesic} $P$ (in $\mathcal H_n$) is a sequence $(x_0,x_1,\ldots,x_n)$
with $x_i \in \mathcal H_n$ for $i=0,1,\ldots,n$, so that there is a permutation 
$(p_0,p_1,\ldots p_n)$ of $(0,1,\ldots n)$ such that the $\ell$th entry of $x_j$
differs from the $\ell$th entry of $x_{j-1}$ if and only if $j=p_\ell$.  
We then say that $P$ \emph{fixed} the $\ell$th entry at time $j$.

We denote by $\inst(f,P)$, for \emph{instability}, the number of colour-jumps
of~$P$ in the colouring $f$, that is, the number of indices~$i$ where
$f(x_i)\neq f(x_{i-1})$. Any such index $i$ shall be called a {\em jump} of
$P$ (in $f$). Let $\inst(f)$ be the maximum value of $\inst(f,P)$ over all
geodesics $P$.

\smallskip

The connection of these concepts and the memoryless consensus problem in distributed
systems is as follows. 
Each point of $\mathcal H_n$ represents a set of $n$ input values (one from each sensor).  
A colouring of~$\mathcal H_n$ corresponds to an assignment of a representative
value for each possible set of input values. 
We prefer colourings that respect the balls of a certain radius as the output value should in some way be representative. 
A geodesic stands for a slowly changing system of inputs (one sensor at a time), and its instability is the number of changes of the representative value. 
We remark that, if one considers arbitrary paths instead of geodesics, there is no bound on the instability as the path might go back and forth between two points with a different output value (see~\cite{DolevR03}).

Now, a colouring that respects $B_t(0^n)$ and $B_t(1^n)$ and has low
instability is a good candidate for a consensus algorithm. 
One is therefore interested in the lowest possible instability.

\begin{problem}[Dolev \& Rajsbaum~\cite{DolevR03}]\label{prob1}
  Given $n\in\mathbb N$, and $t$ valid for $n$, find the minimum value
  $\inst(n,t)$ for $\inst(f)$ over all colourings $f$ of $\mathcal H_n$ that
  respect $B_t(0^n)$ and~$B_t(1^n)$.
\end{problem}

Dolev and Rajsbaum~\cite{DolevR03} prove several special cases: 
$\inst(n,t) \geq 1$ for $n>4t$, $\inst(n,0)=1$, $\inst(n,1)=3$, and
$\inst(2t+1,t)=2t+1$.

In Section~\ref{lobo}, we establish a lower bound of $\lceil \frac{t-1}{n-2t}
\rceil+\lfloor \frac{t-1}{n-2t} \rfloor +3$ on $\inst(n,t)$ (Theorem~\ref{lobo1}) that holds for all values of $n$ and $t$. A similar lower bound also holds for the related parameter
$\winst(n,t)$, which measures the maximum instability of a colouring
considering only a special class of geodesics. This parameter is introduced in
Section~\ref{kdef}.

In Section~\ref{sec:loboonestrip}, we improve our bound to $\inst(2t+2,t)\geq
t+2 + (t+1)\mod 2$ for $t\geq 1$ for the special case of $n=2t+2$
(Theorem~\ref{1faixa}). The basic tool for this result is
Lemma~\ref{l:stronger}, which serves to extend bounds for smaller values of
$t$ to larger values of $t$. This tool is extended in
Section~\ref{sec:lobomorestrips} to arbitrary values of $n$.

In~\cite{DolevR03}, it is also shown that $\inst(n,t) \leq 2t+1$. We conjecture that this bound is indeed the correct value.

\begin{conjecture}[Main conjecture]\label{conj:2t+1}
Let  $n\in\mathbb N$, and $t$ be valid for $n$. Then $\inst(n,t)=2t+1$.
\end{conjecture}


If one can solve Problem~\ref{prob1}, it would be interesting to find all
\emph{optimal} colourings, i.e., colourings for which the bound $\inst(n,t)$ is
attained. In Section~\ref{colo} we exhibit two new classes, $\maj_t(k)$ and
$b_t^k$, of colourings that have instability exactly $2t+1$. Previously only
one such colouring (namely $\maj_t(2t+1)$ in our language) was
known~\cite{DolevR03}.

\section{Candidates for optimal colourings}\label{colo}

We present two classes of colourings that respect the balls $B_t(0^n)$
and $B_t(1^n)$ and have instability $2t+1$.

\subsection{The majority colourings}

For a positive odd value $k$, define $\maj_t(k)$ to be the colouring that
assigns to each point $x\in\mathcal H_n\setminus (B_t(0^n)\cup B_t(1^n))$ the
colour that appears on the majority of the first $k$ entries of~$x$. The balls
$B_t(0^n)$ and $B_t(1^n)$ are coloured canonically with~0 and~1, respectively.

For a positive even value $k$, define the auxiliary class $\maj'_t(k)$ as
the class of colourings $f$ that assign to each point $x$ outside $B_t(0^n)$
and $B_t(1^n)$ (which are coloured canonically) the colour that appears on the
majority of the first~$k$ entries of~$x$, and any colour if both colours
appear equally often in the first~$k$ entries of~$x$. Let $\maj_t(k)\subseteq
\maj'_t(k)$ be the class of those colourings in $\maj'_t(k)$ for
which $f(x) \neq f(y)$ whenever $x$ and $y$ restricted to their first~$k$
entries are the complement of each other. 
In what follows, we often abuse notation and, for a positive even value~$k$, 
write $\maj_t(k)$ for an arbitrary element of $\maj_t(k)$.

\begin{proposition}\label{prop:majo}
  Let $k$, $t$, $n \in \mathbb N$, with $0<k \leq 2t+1\leq n$. Then
  $\inst(\maj_t(k)) = 2t+1$. 
\end{proposition}

The proof of Proposition~\ref{prop:majo} splits into two parts: in
Lemma~\ref{majLeq} we show that no geodesic jumps more than $2t+1$ in
$\maj_t(k)$, and in Lemma~\ref{majGeq} we present a geodesic that
jumps that much. 

Before we turn to these lemmas, let us remark that, when $k > 2t+1$ and odd,
it is easy to find a geodesic that jumps $k$ times in $\maj_t(k)$.  Indeed, we
may start at the point $(01)^{\lfloor n/2\rfloor}0$ and then at each step
switch an entry, from the first to the last. Each of the~$k$ first steps is a
jump. This shows that $\maj_t(k)$, for $k$ large and odd, has instability
larger than $2t+1$.

\medskip

The remainder of this section is devoted to the proof of
Proposition~\ref{prop:majo}, i.e.,~to Lemma~\ref{majLeq} and
Lemma~\ref{majGeq}.  For a geodesic $P=(x_0,x_1,\ldots,x_{n-1},x_n)$, 
the path $Q=(x_n,x_{n-1},\ldots,x_1,x_0)$ is also a geodesic, and is called the
\emph{reverse} of~$P$. Clearly, $\inst(f,P)=\inst(f,Q)$ for any colouring $f$.

\begin{lemma}\label{majLeq}
  If  $0<k \leq 2t+1\leq n$,
  then $\inst(\maj_t(k)) \leq 2t+1$. 
\end{lemma}

\begin{proof}
  Suppose otherwise.  Then there is a geodesic $P=(x_0,x_1,x_2,\ldots,x_n)$ in
  $\mathcal H^n$ with $\inst(\maj_t(k),P)\geq 2t+2$.  Let $m$ be so that the
  $m$th jump of $P$ is the first jump that fixes one of the last $n-k$ entries
  (as $k<2t+2$ there is such an $m$, $1 \leq m \leq n$). Suppose $P$ is chosen such that $m=m(P)$
  is as large as possible. Our plan is to modify $P$ to a geodesic $P'$ with $m(P')>m(P)$, thus obtaining a contradiction.

  Let $i+1$ be the first jump of $P$, and let $\ell$ be the $(2t+2)$nd jump of
  $P$. We assume that $\maj_t(k)(x_\ell)=1$, and thus
  $\maj_t(k)(x_{i})=1$. The other case is analogous.

  As $\ell$ is the $(2t+2)$nd jump of $P$, there are $t+1$ jumps $j$ with
  $j<\ell$ and $\maj_t(k)(x_j)=0$. Hence, $P$ fixed $(t+1)$ $0$'s before time
  $\ell$, and therefore $x_\ell \notin B_t(1^n)$. Thus, since
  $\maj_t(k)(x_\ell)=1$, the majority of the first $k$ entries of~$x_\ell$ is
  not~$0$: it is~$1$ or $k$ is even and $x_\ell$ has as many $0$'s as $1$'s in
  its first~$k$ entries. We can use the same argument on the reverse of $P$ to
  obtain that $x_i\notin B_t(1^n)$, and thus the majority of the first $k$
  entries of $x_i$ is~$1$ or~$k$ is even and~$x_i$ has as many $0$'s as~$1$'s in
  its first~$k$ entries. Thus we showed that
\begin{align}\label{xixell}
 &\textit{ the first $k$ entries of $x_i$ contain at least as many $1$'s as~$0$'s},\\  
 &\textit{ and the same holds for $x_\ell$. }\notag
\end{align}
 
  Because $\maj_t(k)(x_i) = \maj_t(k)(x_\ell) = 1$, the first $k$ entries of
  $x_i$ and of $x_\ell$ are not the complement of each other. So,
  by~\eqref{xixell}, at least one entry within the $k$ first, say the first entry, is~$1$ in both~$x_i$ and~$x_\ell$. This implies that all~$x_j$ with
  $i\leq j \leq \ell$ start with a~$1$.

  Let $S$ be the set of those of the first $k$ entries of $x_i$ that do not
  change in~$P$ between $x_i$ and $x_\ell$. We have just seen that $s:=|S|\geq
  1$. Let $z_1$ be obtained from~$x_i$ by changing the first entry to $0$, and
  for $1< j\leq s$ let $z_j$ be obtained from~$z_{j-1}$ by changing another of
  the entries in $S$. Then 
  
\begin{align}\label{xixell2}
 &\textit{the first $k$ entries of $z_s$ are the complement of the first $k$ entries of $x_\ell$.}
\end{align}

  Let $h$ be the $(2t+1)$st jump of $P$. Then $\maj_t(k)(x_{h-1})=1$ and
  $\maj_t(k)(x_h)=0$. There are $t$ jumps $j\leq h-1$ with
  $\maj_t(k)(x_j)=1$, each fixing a~1 distinct from the first entry. Thus in
  total $x_h$ and $x_{\ell-1}$ have at least $(t{+}1)$ $1$'s, and cannot be in
  $B_t(0^n)$.  In the same way, we see that $x_{i+1}\notin B_t(0^n)$.
  
  Consider $P'=(z_s,z_{s{-}1},\ldots,
  z_1,x_i,x_{i{+}1},\ldots,x_{\ell},y_0,y_1,\ldots,y_{n{-}s{-}\ell{+}i{-}1})$,
  where the $y_j$'s are arbitrarily chosen to complete $P'$ to a
  geodesic. We claim that
 \begin{equation}\label{P'jumps}
 \textit{$P'$ has a jump in its first $s+1$ steps.}
 \end{equation}
  Then we are done because, by the choice of $P$, all the first $m+1$ jumps
  of~$P'$ fix one of the first $k$ entries, contradicting our choice of~$P$.
  
  It remains to prove~\eqref{P'jumps}. As $x_{i+1} \notin B_t(0^n)$ and
  $\maj_t(k)(x_{i+1})=0$, there are at least as many $0$'s as $1$'s among the
  first $k$ entries of $x_{i+1}$. So, since the first entry of~$x_i$ is~$1$,
  but the first entry of $z_1$ is~$0$, there are also at least as many $0$'s
  as $1$'s among the first $k$ entries of $z_1$.
  
  Now, as $x_{i} \notin B_t(1^n)$, also $z_1 \notin B_t(1^n)$. Hence, if the
  first $k$ entries of $z_1$ contain more $0$'s than $1$'s, it follows that
  $\maj_t(k)(z_1)=0$.  As $\maj_t(k)(x_i){=}1$, the geodesic $P'$ has the jump
  $x_i$, which is as desired for~\eqref{P'jumps}.  So we may assume that the
  first $k$ entries of $z_1$ contain exactly as many $0$'s as $1$'s. Thus
  by~\eqref{xixell} and~\eqref{xixell2}, and by the definition of $z_s$, it
  follows that $z_s$ has at least as many $0$'s as $1$'s in its first $k$
  entries, and so at least as many $0$'s as $z_1$ has. Therefore, $z_1
  \notin B_t(1^n)$ implies that $z_s \notin B_t(1^n)$ and hence,
  $\maj_t(k)(z_s)=0$. This finishes the proof of~\eqref{P'jumps}, and thus the
  proof of the lemma.
\end{proof}

\begin{lemma}\label{majGeq}
  If  $0<k \leq 2t+1\leq n$,
  then $\inst(\maj_t(k))\geq 2t+1$. 
\end{lemma}
\begin{proof}
We will prove the following stronger assertion.
\begin{equation}\label{eq:geo}
\begin{minipage}[c]{0.8\textwidth}\em
There exists a $(t+1)$-geodesic $P$ such that
\[
\inst(\maj_t(k),P)\geq 2t+1
\]
and the last point of $P$ is coloured~$0$.
\end{minipage}\ignorespacesafterend 
\end{equation} 

We shall proceed by induction on $k$. Observe that~\eqref{eq:geo} holds for
$k=1$ and for $k=2$. Indeed, for $k=1$, consider the following $(t+1)$-geodesic.
\begin{align*}
P=( & 1^{t+1}0^{n-t-1},         & & [1]\\
    & 1^t00^{n-t-1},            & & [0]\\
    & 1^t010^{n-t-2},           & & [1]\\
    & 1^{t-1}0^210^{n-t-2},     & & [0]\\ 
    & 1^{t-1}0^21^20^{n-t-3},   & & [1]\\
    & 1^{t-2}0^31^20^{n-t-3},   & & [0]\\ 
    & 1^{t-2}0^31^30^{n-t-4},   & & [1]\\
    &\ldots\\
    & 110^{t-1}1^{t-2}0^{n-2t+1}, & & [0]\\
    & 110^{t-1}1^{t-1}0^{n-2t},   & & [1]\\
    & 10^t1^{t-1}0^{n-2t},      & & [0]\\
    & 10^t1^t0^{n-2t-1},        & & [1]\\
    & 0^{t+1}1^t0^{n-2t-1},     & & [0]\\
    & 0^{t+1}1^{t+1}0^{n-2t-2}, & & [0]\\
    & 0^{t+1}1^{t+2}0^{n-2t-3}, & & [0]\\
    & 0^{t+1}1^{t+3}0^{n-2t-4}, & & [0]\\
    & \ldots \\
    & 0^{t+1}1^{n-t-1})         & & [0].
\end{align*}
Note that $P$ jumps $2t+1$ times. For $k=2$, consider either $P$, or the
$(t+1)$-geodesic $P'$ obtained from $P$ by changing the two points
$10^t1^{t-1}0^{n-2t}$ and $10^t1^t0^{n-2t-1}$ to $010^{t-1}1^{t-1}0^{n-2t}$
and $010^{t-1}1^t0^{n-2t-1}$. If $\maj_t(k)(10^t1^{t-1}0^{n-2t}) = 0$, we
choose $P$, otherwise we choose~$P'$.

So suppose we are given a $k \geq 3$. Then $t \geq 1$ and $n \geq 3$. 
Consider $\maj_{t}(k)$ on
\[
\tilde{\mathcal H}^n:=\{x\in\mathcal H^{n}: x(1)=0\text{ and }x(2)=1\},
\]
and observe that this is equivalent to considering $\maj_{t-1}(k-2)$
on $\mathcal H^{n-2}$. Hence, by induction, we know there exists a
$t$-geodesic $\tilde P$ in ${\mathcal H}^{n-2}$ that is as
in~\eqref{eq:geo} for $t-1$. In particular, $\tilde P$ jumps at least
$ 2(t-1)+1=2t-1$ times. Abusing notation slightly, we shall consider
$\tilde P$ as a path in $\tilde{\calH}^{n}$.

Now we extend $\tilde P$ to a geodesic in $\mathcal H^n$ adding two more
jumps. By~\eqref{eq:geo}, we know that $\tilde P$ ends in a point $y$ with
$\maj_t(k)(y)=0$, and with exactly $t+1$ entries equal to $0$ (among these the
first entry). 

Suppose the first point of $\tilde P$, let us call this point $a$, is
coloured~$1$ in $\maj_t(k)$. Then we add the points
$y':=(1,1,y(3),y(4),\ldots)$ and $y'':=(1,0,y(3),y(4),\ldots)$ to the end of
$P'$ and obtain a geodesic $P$ as desired. Indeed, $y'\in B_t(1^n)$ as $y'$
has exactly $t$ $0$'s, hence $\maj_t(k)(y') = 1$, and so we have our first
extra jump. Note that $y''$ has exactly as many $1$'s 
as $y$ (in particular, $y''\notin B_t(1^n)$), and moreover, $y''$ has exactly as many $1$'s in the first $k$ entries as $y$. Thus, $\maj_t(k)(y'') \neq
\maj_t(k)(y)$ only if $k$ is even and $y$ and $y''$ have as many $0$'s as
$1$'s in their first $k$ entries. But in this case, the definition of
$\maj_t(k)$ implies that $0=\maj_t(k)(y'')\neq \maj_t(k)(a)=1$. Therefore,
$\maj_t(k)(y'')=0$, and we have the second extra jump, implying that $P$ is as
in~\eqref{eq:geo}.

It remains to analyse the case where $\maj_t(k)(a)=0$. In this case,
note that as~$\tilde P$ starts and finishes with colour~$0$, it jumps
an even number of times, that is, at least~$2(t-1)+2=2t$ times. Thus
we need to add only one more jump. If we build $P$ in the same way as
above, $P$ jumps at least $2t+1$ times, but for the same reasons as
above, it ends in a point coloured~$1$. So, instead, let $P$ be
obtained from $\tilde P$ by adding at its beginning the two points
$a'':=(1,0,a(3),a(4),\ldots)$ and $a':=(1,1,a(3),a(4),\ldots)$.
Observe that since $a''$ is the complement of $y$, it has the opposite
colour, i.e.,~$\maj_t(k)(a'')=1$. Hence between $a''$ and $a$ we have
at least one jump. So $P$ is a well-ending $(t+1)$-geodesic with at
least $2t+1$ jumps, as desired.
\end{proof}

\subsection{The partition colourings}

We present a second class of colourings, the colourings $b_t^k$, which respect the balls $B_t(0^n)$
and $B_t(1^n)$ and have instability $2t+1$. Before that, we define the auxiliary colouring $a^{\calQ}_j$ that will be used in the definition of $b_t^k$. 

Let $m$, $s$ and~$t$ be such that $m \geq (s+1)(t+1)$. Let $\calQ$ be a
partition of $[m]$ into $s+1$ sets of size at least $t+1$ each.  For $j=0$,
$1$, let $a^{\calQ}_j$ be the following colouring of $\calH_n$.  Let
$a^{\calQ}_j(x)=j$ if and only if, in at least one of the sets in $\calQ$, all
entries are~$j$. It is not difficult to see that $a^{\calQ}_j$ respects both
$B_s(j^m)$ and $B_t((1-j)^m)$.

Consider a geodesic $P=(x_0,x_1,\ldots,x_m)$ in $\calH_m$. Note
that, if $i$ is a jump of $P$ in $a^{\calQ}_j$, then for some set $Q$
in $\calQ$ we have that $x_{\ell}(q) = j$ for all $q \in Q$ either for
$\ell = j-1$ or for $\ell = j$, but not for both. We say that the jump $i$ is
associated to this set $Q$. Thus there are at most two jumps in $P$
associated to the same set $Q$ in $\calQ$. This implies that
$a^{\calQ}_j$ jumps at most $2|\calQ|=2(s+1)$ times.

Now, let $k$, $s$, $t$, and $n$ be such that $k$ is odd, $s \geq -1$, 
$t = s + (k+1)/2$, and $n \geq (s+1)(t+1)+k$. 
Note that $k \leq 2t+1$ because $s \geq -1$. 
Let $\calQ$ be a partition of $[n-k]$ into $s+1$ sets of size at least $t+1$ each. 
(If $s = -1$, then $n=k$ and $\calQ = \emptyset$.)
We shall define the colouring $b_t^k = b_t^k(\calQ)$ using $a^{\calQ}_0$ and $a^{\calQ}_1$. 
We abuse notation and assume that $a^{\calQ}_j(y)=1-j$ if $\calQ$ or $y$ is empty.

For each point $x$, if the majority of the first $k$ entries of $x$ is 1, 
then let $b_t^k(x)=a^{\calQ}_0(x')$, where $x'$ is $x$ without the first $k$ entries. 
If the majority of the first $k$ entries of $x$ is 0, then let $b_t^k(x)=a^{\calQ}_1(x')$.  
In both cases, we sometimes abuse notation and write that $b_t^k = a^{\calQ}_j$ in $x$.

It is not difficult to see that $b_t^k$ respects the balls $B_t(0^n)$ and $B_t(1^n)$. 
Indeed, let us suppose the majority of the first~$k$ entries of some point $x$ is~$1$, 
and hence $b_t^k = a^{\calQ}_0$ (the other case is symmetric). 
If $x$ has at most $t$ entries equal to $0$, clearly no set in $\mathcal Q$ can only consist of $0$'s, and so $b_t^k(x)=1$. 
On the other hand, if $x$ has at most $t$ $1$'s, then $x'$ has at most $t-(k+1)/2=s$ $1$'s and  
therefore, as $|\calQ|=s+1$, there is a set in $\mathcal Q$ that only consists of $0$'s. Thus $b_t^k(x)=0$ in this case.
Hence, in either case, $b_t^k(x)$ is as desired.

Observe that, for $t=0$ and $k=1$, we have $s=-1$, and hence $n=1$. In this case, $b_0^1 = \maj_0(1)$.

\begin{proposition}\label{prop:block}
  Let $k$, $t$, $n \in \mathbb N$ be such that $k$ is odd, $k \leq 2t+1$ and
  $n \geq (t+1-\frac{k+1}{2})(t+1)+k = \frac{(t+1)(2t+1)-k(t-1)}{2}$. 
  Then $\inst(b_t^k)=2t+1$. 
\end{proposition}

\begin{proof}
  Let $P$ be a geodesic in $\calH_n$. To prove that $P$ jumps at most
  $2t+1$ times in $b^t_k$, first note that at most $k$ jumps of $P$
  are associated to its first~$k$ entries. Second, note that $P$ has at most
  two jumps associated to each set~$Q$ in $\calQ$. Indeed, if $P$ has
  one jump associated to $Q$ while $b_t^k = a^{\calQ}_j$, then~$P$ has
  at most one more jump associated to $Q$ while $b_t^k = a^{\calQ}_{1-j}$. 
  Similarly, if $P$ has two jumps associated to $Q$ while $b_t^k =
  a^{\calQ}_j$, then $P$ has no jumps associated to $Q$ while $b_t^k =
  a^{\calQ}_{1-j}$.

  Also, it is not hard to find a geodesic in $\calH_n$ that jumps
  $2t+1$ times in~$b_t^k$. Consider a point $x_0$ with $(k+1)/2$ 1's
  in the first $k$ entries, and exactly one~$1$ in each of the sets in
  $\calQ$. Then $x_0$ has exactly $t+1$ entries equal to~$1$. Take a
  geodesic that starts in $x_0$, and jumps $k$ times by changing alternatively 
  $1$'s to $0$'s and $0$'s to $1$'s within the first $k$ entries. After that, 
  we have that $b_t^k = a^{\calQ}_1$. So we can jump twice per set $Q$ in
  $\calQ$ by changing all entries in $Q$ to~$1$ first, and then changing
  the unique entry in $Q$ that started with a $1$ to a~$0$.
\end{proof}

\section{Well-ending geodesics and $k$-defined colourings}\label{kdef}

A geodesic in $\mathcal H_n$ is called an \emph{$m$-geodesic} if it starts in
a point of $\mathcal H_n$ which has exactly $m$ entries equal to $1$. (It
then ends in a point which has exactly $m$ entries that equal $0$.)

Let $f$ be a colouring of $\mathcal H_n$ and $t_f$ be the maximum $t$ such
that $f$ respects $B_t(0^n)$ and $B_t(1^n)$. (If $f(0^n)=1$ or $f(1^n)=0$,
then set $t_f=-1$.)

If $P$ is a geodesic whose first point is coloured~$1$ in $f$, or
whose last point is coloured~$0$ in $f$, we say $P$ {\em ends
  well} (in $f$). Let $\winst(f)$ denote the maximum value of $\inst(f,P)$,
taken over all well-ending $(t_f+1)$-geodesics~$P$. In analogy to
Problem~\ref{prob1}, we ask the following.

\begin{problem}\label{prob:neu}
Given $t$ valid for $n$, which is the smallest value $\winst(n,t)$ such that
$\winst(n,t)=\winst(f)$ for some colouring $f$ with $t_f=t$?
\end{problem}

Observe that $\winst(f)\leq \inst(f)$ for every colouring $f$. 
Moreover,  $\min_{t\leq s\leq (n-1)/2}\{\winst(n,s)\}\leq \inst(n,t)$ for all $t$ valid for $n$. 

Call a colouring $f$ \emph{$k$-defined}\footnote{ We remark that
in~\cite[pg.\ 39]{DolevR03},  \emph{one-bit defined}
colourings are introduced. This definition differs from ours (for $k=1$) as we
canonically colour the balls $B_t(0^n)$ and $B_t(1^n)$. For instance,
$maj_t(1)$ is $1$-defined, but not one-bit defined.} if there are $k$ indices
such that $f(x)=f(y)$ for any two points $x$, $y \in\mathcal H_n\setminus
(B_t(0^n) \cup B_t(1^n))$ that coincide in all entries given by these $k$
indices. 
A $k$-defined colouring that is not $(k{-}1)$-defined is called {\em
  strictly $k$-defined}.  For instance, $\maj_t(k)$ is strictly
$k$-defined and $a^{\calQ}_0$ is strictly $n$-defined.

Let $t$ be valid for $n$. For the next lemma, let $F^n(t)$ denote the set of
all strictly $n$-defined colourings $f$ of $\calH_n$ with $t_f=t$, and let 
$F^{<n-2t}(t)$ denote the set of all strictly $k$-defined colourings $f$ of
$\calH_n$ with $0\leq k<n-2t$ and $t_f=t$.

\begin{lemma}\label{lem:neu}
  Let $g:\N\to\N$ be such that $g(s) + 2(t-s) \geq g(t)$ for all $s
  \leq t$.  If $\winst(f') \geq g(t')$ for all $t'$ valid for $n$ and all $f' \in F^n(t')$, then
  $\winst(f) \geq \min\{g(t),2t+2\}$ for all $t$ 
  valid for $n$ and  all $f \in F^{<n-2t}(t)$.
\end{lemma}

This lemma could be used as a step towards a solution of Problem~Ê\ref{prob1}.
In fact, consider $g(s) = 2s+1$ and note that such $g$ satisfies the
assumption of the lemma. If we could prove that $\winst(f) \geq 2t_f+1$ for all
colourings $f$ that are strictly $k$-defined with $k \geq n-2t_f$, then
Lemma~\ref{lem:neu} would assure this bound holds for all colourings $f$ of
$\calH_n$, and thus imply Conjecture~\ref{conj:2t+1}.

The rest of this section is devoted to the proof of Lemma~\ref{lem:neu}.

\begin{proof}[Proof of Lemma~\ref{lem:neu}]
  Let $t$ be valid for $n$, and let $f \in F^{<n-2t}(t)$. 
  We assume that the defining entries of $f$ are the first $k < n-2t$. 
  Our aim is to find a well-ending $(t+1)$-geodesic~$P$ for $f$ 
  that jumps at least $\min\{g(t),2t+2\}$.

  Consider the hypercube $\mathcal H_k$, and let $f'$ be the colouring
  of~$\mathcal H_k$ that assigns to each $x$ in $\mathcal H_k$ the colour
  that~$f$ assigns to the point of~$\mathcal H_n$ obtained from~$x$ by 
  adding $(t{+}1)$ $1$'s and $(n{-}k{-}t{-}1)$ $0$'s at the end.  
  Observe that
  \begin{equation}\label{eq:f''}
  f'\text{ is strictly }k\text{-defined,}
  \end{equation} 
  as $f$ is.
  Now if $f'(0^k)=f(0^k1^{t+1}0^{n-k-t-1})=1$, then there exists a
  $(t{+}1)$-geodesic in $\mathcal H_n$ that jumps at least $2t+2$ times. 
  For example, consider
\begin{align*}
Q=( 
& 0^k1^{t+1}0^{n-k-t-1}, & & [1] \\
& 0^k1^{t}0^{n-k-t}, & & [0] \\ 
& 0^k1^{t}0^{n-k-t-1}1, & & [1]\\
& 0^k1^{t-1}0^{n-k-t}1, & & [0]\\
&\ldots\\
& 0^k10^{n-k-t-1}1^t, & & [1]\\
& 0^k0^{n-k-t}1^t, & & [0]\\
& 0^k0^{n-k-t-1}1^{t+1}, & & [1]\\
& 0^k0^{n-k-t-2}1^{t+2}, & & [1]\\
& 0^k0^{n-k-t-3}1^{t+3}, & & [1]\\ & \ldots \\
& 0^k0^{t+1}1^{n-k-t-1}, & & [1]\\
& 10^{k-1}0^{t+1}1^{n-k-t-1}, & & [?]\\ &\ldots\\
& 1^k0^{t+1}1^{n-k-t-1}) & & [?].
\end{align*}
Similarly, if $f'(1^k) = 0$, consider the geodesic obtained from $Q$ by
swapping all $0$'s and $1$'s. Its reverse is a $(t+1)$-geodesic that jumps at
least $2t+2$ times.
%

Therefore, we assume from now on that $f'(0^k)=0$ and $f'(1^k) = 1$, in
other words, that $f'$ respects $B_0(0^k)$ and $B_0(1^k)$ and thus $t_{f'} \geq 0$. 
By~\eqref{eq:f''}, we can use the assumption of the lemma for $s:=t_{f'}$ and $f'$ to
obtain an $(s{+}1)$-geodesic $P'=(p_0,p_1,\ldots,p_k)$ in $\mathcal H_k$ such
that $\inst(f',P')\geq g(s)$. Furthermore $P'$ starts with colour~$1$, or
ends with colour~$0$, say the former (the other case is symmetric). Note that
we can adjust $P'$ without decreasing its instability so that, when $P'$ is
inside one of the balls, it has exactly $s$ $0$'s or $1$'s, respectively. That
is, we may assume each point in $P'$ has at least $s$ $0$'s and at least $s$
$1$'s.

We shall now add a few $0$'s and $1$'s to each point in $P'$ in order to
make it a path $P''$ in $\mathcal H_n$.
Then, we shall extend $P''$ to a $(t+1)$-geodesic $P$ in $\mathcal H_n$, and
make it jump $2(t-s)$ additional times at the border of one of the balls
$B_t(0^n)$ or $B_t(1^n)$. As we explain ahead, these two goals are achieved by
the geodesic
\begin{align*}
  P=(
  & p_01^{t-s}0^{n-k-t+s}, & & [1] \\
  & p_01^{t-s-1}0^{n-k-t+s+1}, & & [0] \\
  & p_01^{t-s-1}0^{n-k-t+s}1, & & [1]\\
  & p_01^{t-s-2}0^{n-k-t+s+1}1, & & [0]\\
  & p_01^{t-s-2}0^{n-k-t+s}1^2, & & [1]\\
  &\ldots\\
  & p_010^{n-k-t+s+1}1^{t-s-2}, & & [0]\\
  & p_010^{n-k-t+s}1^{t-s-1}, & & [1]\\
  & p_00^{n-k-t+s+1}1^{t-s-1}, & & [0]\\
  & p_00^{n-k-t+s}1^{t-s}, & & [1]\\
  & p_10^{n-k-t+s}1^{t-s}, & & [?]\\
  & p_20^{n-k-t+s}1^{t-s}, & & [?]\\
  & \ldots \\
  & p_k0^{n-k-t+s}1^{t-s}, & & [?]\\
  & p_k0^{n-k-t+s-1}1^{t-s+1}, & & [?]\\
  & p_k0^{n-k-t+s-2}1^{t-s+2}, & & [?]\\ & \ldots \\
  & p_k0^{t-s}1^{n-k-t+s}) & & [?].
\end{align*}

The initial part of $P$ jumps $2(t-s)$ times, since $p_0$ has exactly $s+1$
entries equal to $1$. Moreover, the last part of $P$ (where the colours are
marked as ``?'') jumps at least $g(s)$ times. Indeed, as $p_0,\ldots,p_k$ have
at least $s$ $0$'s and at least $s$ $1$'s, the points $p_i0^{n-k-t+s}1^{t-s}$,
for $i=0,\ldots,k$, have at least $n-k-t+s+s=n-k-t+2s > t+2s \geq t$ $0$'s and
at least $t$ $1$'s, so, this part of $P$ enters the balls exactly when $P'$
does.  Thus the part of~$P$ that goes through the points
$p_i0^{n-k-t+s}1^{t-s}$ for $i=0,\ldots,k$ jumps exactly when~$P'$ does, that
is, at least $g(s)$ times. So, by our assumption on $g$, it follows that $P$
jumps at least $2(t-s)+g(s)\geq g(t)$ times, completing the proof of the
lemma.
\end{proof}

\section{Lower bounds on $\inst(n,t)$ and $\winst(n,t)$}\label{lobo}

\subsection{The zig-zag bound}

In this section we prove lower bounds for $\inst(n,t)$ and $\winst(n,t)$. Recall that any lower bound on $\winst(f)$ also serves as a lower bound for $\inst(f)$. We start with a bound for all values of $n$ and valid $t$, which we obtain from a fairly basic zig-zag argument.

Later, in Theorem~\ref{1faixa} and Proposition~\ref{morestrips}, the bounds from Theorem~\ref{lobo1} will be improved for the special cases $n=2t+2$ and $n=2t+3$.

\begin{theorem}[The zig-zag bound]\label{lobo1} Let $n \in \mathbb N$ and let $t \geq 0$ be valid for $n$. 
Then
\begin{enumerate}
\item[(a)] $\winst(n,t)\geq \lfloor \frac{t}{n-2t} \rfloor + \lceil \frac{t}{n-2t} \rceil+1$,
\item[(b)] $\inst(n,t)\geq \lfloor \frac{t-1}{n-2t} \rfloor +\lceil \frac{t-1}{n-2t} \rceil + 3$,\  if\  $t\geq 1$. \label{lobo1b}
\end{enumerate}
\end{theorem}


We remark that Theorem~\ref{lobo1} (a) proves Conjecture~\ref{conj:2t+1} for $t=0$ 
and Theorem~\ref{lobo1} (b) proves Conjecture~\ref{conj:2t+1} for $t=1$. 
This has been shown earlier in~\cite{DolevR03}.

We dedicate the rest of this subsection to the proof of Theorem~\ref{lobo1}.

\begin{proof}[Proof of Theorem~\ref{lobo1}]
  Let $f$ be a colouring with $t_f \geq 0$ and let $t=t_f$. For~(a), 
  our aim is to find a well-ending $(t+1)$-geodesic $P$ that jumps at least
  $\lfloor \frac{t}{n-2t} \rfloor + \lceil \frac{t}{n-2t} \rceil+1$ times in $f$.
  
  As $t=t_f$, there is a point $x\in\mathcal H^n$ that has exactly $(t+1)$
  $1$'s or $(t+1)$ $0$'s, and that is coloured $1$ or $0$, respectively. Say
  the former holds for~$x$ (the other case is symmetric).
 
  We let $P$ start in $x$, then enter $B_t(0^n)$, then go to $B_t(1^n)$, come
  back to $B_t(0^n)$, go to $B_t(1^n)$ again, etc., until $P$ has used up all
  of its entries. For example, if $x=1^{t+1}0^{n-t-1}$, we let $P$ pass next
           through $1^{t}0^{n-t}$                 
  and then through $1^ t0^t1^{n-2t}$,             
           through $1^{t-(n-2t)}0^{n-t}1^{n-2t}$, 
           through $1^{t-(n-2t)}0^t1^{2n-4t}$,    
  and so on.
  
  We can do this until one of the following two things happens. Firstly,
  coming from $B_t(0^n)$, we might end in the complement of $x$ with $(t+1)$
  $0$'s (just before reaching $B_t(1^n)$). This will happen exactly when
  $n=\ell (n-2t)$ for some odd $\ell$, which is the case if and only if $n-2t$
  divides $t$. Then we will have jumped at least $\ell$ times and
$$ \ell \ = \ \frac{n}{n-2t} \ = \ \frac{2t}{n-2t} + 1 
\ = \ \left\lfloor \frac{t}{n-2t} \right\rfloor + \left\lceil \frac{t}{n-2t} \right\rceil + 1. $$

Secondly, on our way from $B_t(1^n)$ to  $B_t(0^n)$, we might reach a point 
of $\mathcal H_n \setminus (B_t(0^n)\cup B_t(1^n))$ which has no more unused $1$'s. 
This happens if and only if $n-2t$
does not divide $t$. Then we have to return in the direction of $B_t(1^n)$ to 
end in the complement of $x$ (if we are not already there). In this case, 
we have jumped at least 
$$ 1 + 2\cdot \left\lfloor \frac{n-(n-2t)}{2(n-2t)} \right\rfloor + 1
      \ = \ 2\cdot \left\lfloor \frac{t}{n-2t} \right\rfloor + 2 
      \ = \ \left\lfloor \frac{t}{n-2t} \right\rfloor + \left\lceil \frac{t}{n-2t} \right\rceil + 1 $$ 
times, because at least one jump is achieved during the first $n-2t$ steps, 
then we get at least $2$ jumps for every $2(n-2t)$ steps, 
and finally we jump at least once more in the last part of $P$ when going through  $B_t(1^n)$.
Note that, by the construction of $P$, we have to end up in one of the two situations just described. 
This completes the proof of (a).



For (b), the proof is similar, the difference being that we let $P$ start inside $B_t(0^n)$, have $x$ as its second point, then re-enter $B_t(0^n)$, and then go on in a zig-zag fashion as before. 
We will obtain $2$ jumps in the beginning,  at least one jump during the next $n-2t$ steps of $P$, and then $2$ jumps every $2(n-2t)$ steps. Finally we might ensure another jump depending on whether $n-2=\ell (n-2t)$ for some odd $\ell$ or not. More precisely, if $n-2=\ell (n-2t)$ for some odd $\ell$, that is, if $n-2t$ divides $t-1$, then we get 
\[
\ell + 2 =\frac{n-2}{n-2t}+2
\ = \ \left\lfloor \frac{t-1}{n-2t} \right\rfloor + \left\lceil \frac{t-1}{n-2t} \right\rceil + 3
\]
jumps, and otherwise, we also get
\[
2+1+2\cdot \lfloor \frac{n-2-(n-2t)}{2(n-2t)} \rfloor +1\ = \ \left\lfloor \frac{t-1}{n-2t} \right\rfloor + \left\lceil \frac{t-1}{n-2t} \right\rceil + 3
\]
jumps, which is as desired. 
Clearly, we need here that $t\geq 1$, because otherwise we could not enter $B_t(0^n)$ twice in the beginning.
\end{proof}

\subsection{Better bounds for  one strip}\label{sec:loboonestrip}

In this subsection we will concentrate on the case when $\mathcal H_n$
contains, besides the balls, only one `strip' of points which all have the
same number of entries equal to $0$ and equal to $1$. 
That is, we treat the case $n=2t+2$.

From Theorem~\ref{lobo1}, we have that 
$\winst(2t+2,t) \geq t+1$ and 
$\inst(2t{+}2,t) \geq t+2$ for $t \geq 1$. 
The following result improves this bound. 

\begin{theorem}\label{1faixa}
$\inst(2t+2,t) \geq \winst(2t+2,t) \geq t+3$ for all $t \geq 2$. 
\end{theorem}


We will prove Theorem~\ref{1faixa} by combining the next two lemmas. 
The first of these is a tool for extending bounds for small values 
of $t$ to larger values of $t$.

\begin{lemma}\label{l:stronger}
Let $y_0$, $t_0$ and $t \in \mathbb N$ with $t \geq t_0$. 
If $\winst(2t_0+2,t_0) \geq y_0$ for some $t_0 \geq 0$ 
then $\winst(2t+2,t)\geq y_0 + t - t_0$.
\end{lemma}

\begin{proof}
  We proceed by induction on $t$. 
  The base, for $t=t_0$, follows directly from
  the hypothesis of the lemma. For $t>t_0$, consider a colouring $f$ with
  $t_f=t$ of the hypercube $\mathcal H_n$ of dimension $n=2t+2$. 

  Define a colouring $g$ of the hypercube $\mathcal H_{n-2}$ by assigning to
  each $x'$ in $\mathcal H_{n-2}$ the value $g(x')=f(01x')$. Then $g$ is such
  that $t_g=t-1$. Indeed, any point of $\mathcal H_{n-2} \setminus
  (B_{t-1}(0^{n-2})\cup B_{t-1}(1^{n-2}))$ is a witness to this. We may thus
  apply the induction hypothesis to obtain a well-ending $t$-geodesic $\tilde
  P$ in $\mathcal H_{n-2}$ that jumps at least $y_0+t-1-t_0$ times in
  $g$. Extending each point $x'$ of $\tilde P$ to the point $01x'$ of
  $\mathcal H_n$, we obtain a path $P'$ in $\mathcal H_n$ that jumps at least
  $y_0+t-1-t_0$ times in $f$.

  Let $01a$ and $01z$ be the first and last point of $P'$ respectively. 
  Note that since $\tilde P$ is well-ending, either $g(01a)=1$ or $g(01z)=0$ 
  (or both). 
  We extend $P'$ to $P$ by adding to its beginning the points $00a$ and
  $10a$, if $g(01a)=1$, and the points $11z$ and $10z$ to its end
  otherwise. As we thus pass once more through either $B_t(0^n)$ or
  $B_t(1^n)$, our extension $P$ of $P'$ jumps at least once more than $P'$, 
  that is, $y_0+t-t_0$ times in total. Clearly, $P$ is a $(t+1)$-geodesic, 
  and so is its reverse, because $n=2t+2$. Now at least one of the two, 
  $P$ or its reverse, has to be well-ending, which completes the proof of
  the lemma.
%
 \end{proof}
 
The next lemma takes care of the base case $t=t_0$ for Lemma~\ref{l:stronger}.
It also confirms Conjecture~\ref{conj:2t+1} for $n=2t+2$ and small values of~$t$.


\begin{lemma}\label{l:base} 
  $\winst(2t+2,t) \geq 2t+1$ for $t=0,1,2$. 
\end{lemma}

\begin{proof}
  The case $t=0$ is trivial. For $t=1$, let $f$ be a colouring of $\calH_4$
  such that $t_f=1$. Note that there are two points $x$ and $y$ in $\calH_4$
  with exactly $t+1=2$ entries equal to $1$, differing in exactly two entries
  (that is, such that $||x-y||^2 = 2$), and such that $f(x)=f(y)$.  For
  example, two of the three points $1100$, $1010$, $1001$ must have the same
  colour in $f$. Now it is easy to construct a well-ending $2$-geodesic
  that starts in $x$ and jumps at least three times.

  For $t=2$, let $f$ be a colouring of $\calH_6$ such that $t_f=2$.  Observe
  that we only need to find three points $x$, $y$, $z$, all with exactly
  $t+1=3$ entries equal to~$1$, such that $||x-y||^2=||y-z||^2=2$,
  $||x-z||^2=4$, and $f(x)=f(y)=f(z)$. Indeed, if we have such points, it is
  easy to construct a well-ending $3$-geodesic that starts in $x$ and
  jumps at least five times.
  
  The proof of the existence of $x$, $y$ and $z$ is a case analysis. By
  rearranging the order of the entries, we may assume the points $x=111000$
  and $y=110100$ have the same colour $j$ in $f$. If one among $x'=100110$,
  $y'=100101$ and $z'=010101$ has colour $j$, then we may take it as our third
  point~$z$. If not, then $x'$, $y'$ and $z'$ all have colour $1-j$ and form a
  triple of points as desired.
\end{proof}

\begin{proof}[Proof of Theorem~\ref{1faixa}]
  The statement is an immediate consequence of
  Lemma~\ref{l:stronger} and Lemma~\ref{l:base} for $t=2$. 
\end{proof}

\subsection{The extension method for more strips}\label{sec:lobomorestrips}

We now extend the results from the previous subsection to the general case,
when we have more `strips'. The main result of this subsection,
Proposition~\ref{morestrips}, is an extension of Lemma~\ref{l:stronger}.
%
%

\begin{proposition}\label{morestrips}
  Let $n$, $y_0, t_0\in\mathbb N$ and let $t \geq t_0$ be valid for $n$ and
  such that $n-2t$ divides $t-t_0$. 
If $\winst(n,t_0) \geq y_0$, then $\winst(n,t)\geq y_0 + 2\frac{t - t_0}{n-2t}$.
\end{proposition}

Clearly, Proposition~\ref{morestrips} can be used in the same way as
Lemma~\ref{l:stronger} to improve Theorem~\ref{lobo1}.  The next lemma takes
care of the base case $t=t_0$ for Proposition~\ref{morestrips}, for the case
$n=2t+3$. It also confirms Conjecture~\ref{conj:2t+1} for $n=5$ and~$t=1$.


\begin{lemma}\label{l:base2} 
  $\winst(5,1) \geq 3$, $\winst(7,2) \geq 4$, and $\winst(9,3) \geq 4$. 
\end{lemma}

\begin{proof}
  We start proving that $\winst(5,1) \geq 3$.
  Let $f$ be a colouring of $\calH_5$ with $t_f=1$. We say a point $x$ in
  $\calH_5$ is \emph{good} (in $f$) if  there is a $j\in\{0,1\}$ so that $x$ has exactly two entries equal to~$j$
  and $f(x)=j$. Also, we say that two points $x$ and $y$ in $\calH_5$ are
  \emph{neighbours} if $||x-y||^2 = 2$ and they have the same number of
  entries equal to~1.

  First of all, we observe that, if there are two good points $x$ and $y$ that
  are neighbours, then it is easy to construct a well-ending 2-geodesic that
  jumps the required number of times (in the same way as in Lemma~\ref{l:base}). 
  So we may assume that
  \begin{equation}\label{noxy}
  \text{if $x$ and $y$ are good in $f$, then they are not neighbours.}
  \end{equation}

  Second, we may assume that 
  \begin{equation}\label{oppo}
    \text{if $x$ is good in $f$ then its complement is not good in $f$.}
  \end{equation}
  Indeed, if a point $x$ and its complement are good in $f$, then we may
  obtain a well-ending 2-geodesic as desired by starting out at $x$, going to
  $B_1(j^5)$, then going to $B_1((1-j)^5)$, and then ending at the complement
  of $x$.

  As $t_f=1$, there is a point $w$ that is good in $f$. By symmetry, we can assume
  that $w=00011$. Now, because of~\eqref{noxy}, at most one of the points
  $11000$, $10100$ and $01100$ is good. So, at least one of them, say $11000$,
  has colour $0$ in $f$.
  Consider the 2-geodesic
  $$(00011 [1], 00001 [0], 01001 [0], 11001 [?], 11000 [0], 11100 [1]).$$
  Its third point has colour~0 because of~\eqref{noxy} and its last point has
  colour $1$ because of~\eqref{oppo}. So this well-ending 2-geodesic only jumps
  less than three times if $f(11001)=0$. But in this case, we use~\eqref{noxy}
  to see that $f(11010)=1$, and consider the well-ending 2-geodesic 
  $$(00011 [1], 00010 [0], 10010 [0], 11010 [1], 11000 [0], 11100 [1]),$$ 
  that jumps $4>3$ times. (Again, $f(10010)=0$ because of~\eqref{noxy}.)
  This concludes the proof that $\winst(5,1) \geq 3$.

  The idea for the other cases is similar to the one used in the proof 
  of Lemma~\ref{l:stronger}. We reduce the problem to 5 entries, obtaining 
  as above a `partial' geodesic that jumps at least three times, and extend 
  it so that it jumps least four times, as needed. 

  For $\winst(7,2)$, let $f$ be a colouring of $\calH_7$ with $t_f=2$. 
  Let $w$ be a point in $\calH_7$ with exactly 3 entries equal to $j$ and
  such that $f(w)=j$. 
  By symmetry, we may assume that the first two entries of $w$ are $01$. 

  Define a colouring $g$ of the hypercube $\mathcal H_5$ by assigning to
  each $x'$ in $\mathcal H_5$ the value $g(x')=f(01x')$. Then $g$ is such
  that $t_g=1$. Indeed, the point $w'$ in $\calH_5$ such that $w=01w'$
  serves as a witness to this. 

  As $\winst(5,1) \geq 3$, there is a well-ending 2-geodesic $\tilde P$ in
  $\calH_5$ that jumps at least three times in $g$.  Extending each point $x'$
  of $\tilde P$ to the point $01x'$ of $\calH_7$, we obtain a path $P'$ in
  $\calH_7$ that jumps at least three times in $f$.  If~$\tilde P$ jumps
  exactly three times, then it ends well in both of its ends. Thus we can
  extend $P'$ in one of its ends, passing by the neighbouring ball, so that it
  jumps once more, and the result will be a well-ending 3-geodesic as
  desired. If, on the other hand, $\tilde P$ jumps at least four times, then
  we just extend it in any way
  so that the resulting geodesic is still well-ending. This completes the 
  proof that $\winst(7,2) \geq 4$.
  The proof that $\winst(9,3) \geq 4$ is similar, so we omit it.
\end{proof}

\begin{corollary}\label{2faixasnovo}
Let  $t \geq 1$. Then
$$\winst(2t+3,t) \ \geq \ 2 + \frac{2t + (t\mod 3)}{3}.$$
\end{corollary}

\begin{proof}
  We obtain the bound by applying Proposition~\ref{morestrips} to $n=2t+3$ and
  the base cases obtained from Lemma~\ref{l:base2}: $t_0=1$ with $y_0=3$, 
  $t_0=2$ with $y_0=4$, and $t_0=3$ with $y_0=4$. 
\end{proof}

This bound improves by one the bound from Theorem~\ref{lobo1} (a) for 
$n=2t+3$ and $t \mod 3 = 0$ or $1$, and by two for $t \mod 3 = 2$.

The rest of this section is dedicated to the proof of Proposition~\ref{morestrips}.

\begin{proof}[Proof of Proposition~\ref{morestrips}]
  We proceed by induction on $i=i(n,t):=\frac{t-t_0}{n-2t}$. 
  The base, for $i=0$ (i.e.,~$t=t_0$), follows directly from the hypothesis of
  the lemma. For $i>0$, consider a colouring $f$ of the hypercube $\mathcal
  H_n$ with $t_f=t$.
    
  By the definition of $t_f$, there is an $x$ in $\calH_n$ with exactly $t+1$
  entries equal to $f(x)$. As $t$ is valid for $n$, we know that $x$ has at
  least $t$ entries equal to $1-f(x)$. So, as $n-2t \leq t-t_0 \leq t$, we may
  assume that $x=0^{n-2t}1^{n-2t} x'$, where $x'\in \mathcal H_{n'}$ for
  $n':=n-2(n-2t)$.

  Define a colouring $g$ of the hypercube $\mathcal H_{n'}$ by assigning to
  each $x''$ in $\mathcal H_{n'}$ the value $g(x'')=f(0^{n-2t}1^{n-2t} x'')$.
  Then $g$ is such that $t_g=t':=t-(n-2t)$. Indeed, $g$ respects the balls
  $B_{t'}(0^{n'})$ and $B_{t'}(1^{n'})$ because $f$ respects the balls
  $B_t(0^n)$ and $B_t(1^n)$, and the point $x'$ has exactly $t-(n-2t)+1=t'+1$
  entries equal to $g(x') = f(x)$.
  
  Note that $t'$ is valid for $n'$ and that $n'-2t'=n-2t$ divides $t'-t_0$. Moreover,
  \[
  i(n',t')=\frac{t-t_0-(n-2t)}{n-2t}=i(n,t)-1.
  \]
  So, we may apply the induction hypothesis to $\mathcal H_{n'}$ and $g$ to
  obtain a well-ending $(t'+1)$-geodesic $\tilde P$ in $\mathcal H_{n'}$ that
  jumps at least $y_0+2\frac{t-t_0}{n-2t}-2$ times in $g$.  We suppose that
  the first point $\tilde a$ of $\tilde P$ is such that $g(\tilde a)=1$. In
  other words, we suppose that $\tilde P$ ends well in its first point. The
  other case is analogous.

  Extending each point $x''$ of $\tilde P$ to the point $0^{n-2t}1^{n-2t}x''$
  of $\mathcal H_n$, we obtain a path $P'$ in $\mathcal H_n$ that jumps at
  least $y_0+2\frac{t-t_0}{n-2t}-2$ times in~$f$.
  Let $z=0^{n-2t}1^{n-2t}z'$ be the last point of $P'$. 
  If $f(z)=0$, then we extend $P'$ to $P$ by adding to its end the points \\ \vspace{-2mm} \ \\
  $0^{n-2t-1}1^{n-2t+1}z'$, \\ 
  $0^{n-2t-1}101^{n-2t-1}z'$, \\ 
  $0^{n-2t-1}10^21^{n-2t-2}z'$, \\ $\ldots$ \\
  $0^{n-2t-1}10^{n-2t}z'$, \\  
  $10^{n-2t-2}10^{n-2t}z'$,\\ 
  $1^20^{n-2t-3}10^{n-2t}z'$, \\ $\ldots$ \\ 
  $1^{n-2t}0^{n-2t}z'$. \\ 

  As we thus pass once through $B_t(1^n)$, and then through $B_t(0^n)$, 
  our geodesic $P$ jumps at least two times more than $P'$.
  
  On the other hand, if $f(z)=1$, 
  then we extend $P'$ to $P$ by adding to its end the points \\ \vspace{-2mm} \ \\
  $0^{n-2t+1}1^{n-2t-1}z'$, \\ 
  $0^{n-2t+2}1^{n-2t-2}z'$, \\
  $0^{n-2t+3}1^{n-2t-3}z'$, \\ $\ldots$ \\
  $0^{2n-4t-1}1z'$, \\  
  $10^{2n-4t-2}1z'$,\\ 
  $1^20^{2n-4t-3}1z'$, \\ $\ldots$ \\ 
  $1^{n-2t}0^{n-2t-1}1z'$, \\ 
  $1^{n-2t}0^{n-2t}z'$. \\  
    
  We passed once through  $B_t(0^n)$, and then through $B_t(1^n)$, 
  thus again our geodesic $P$ has at least two more jumps than $P'$. 
  
  So, in either case, $P$ jumps at least $y_0+2\frac{t-t_0}{n-2t}$ 
  times in total. By construction, $P$ is a well-ending $(t+1)$-geodesic, as desired.
 \end{proof}
%
%
%
%
%


\begin{thebibliography}{10}\label{bibliography}
\bibitem{EATCS06} Aspnes, J., Busch, C., Dolev, S., Fatourou, P., Georgiou, C., Shvartsman, A., 
  Spirakis, P., and Wattenhofer, R., \emph{Eight open problems in distributed computing},
  Bulletin of the European Association for Theoretical Computer Science,
  Distributed Computing Column, 90:109--126, October 2006.  
\bibitem{BeckerRRR10} Becker, F., Rajsbaum, S., Rapaport, I., and R\'emila,
  E., \emph{Average long-lived binary consensus: Quantifying the stabilizing role
  played by memory}, Theoretical Computer Science, Volume 411, Issues 14--15, pp.1558--1566, 2010.
\bibitem{DavidovitchDR07} Davidovitch, L., Dolev, S., and Rajsbaum, S., 
  \emph{Stability of multivalued continuous consensus},  
  SIAM Journal on Computing, Vol.\ 37 , Issue 4, pp.1057--1076, 2007.
\bibitem{DolevH08} Dolev, S., and Hoch, E.N., 
  \emph{OCD: obsessive consensus disorder (or repetitive consensus)}, 
  PODC'08 Proceedings of the Twenty-Seventh ACM Symposium on Principles
  of Distributed Computing, 356--404, 2008.
\bibitem{DolevR03} Dolev, S., and Rajsbaum, S., \emph{Stability of long-lived consensus}, 
  Journal of Computer and System Sciences, Vol. 67, Num. 1, August 2003, pp.26--45.
\bibitem{RapaportR10} Rapaport, I., and R\'emila, E., 
  \emph{Average long-lived memoryless consensus: the three-value case}, 
  Lecture Notes in Computer Science, 2010, Volume 6058, 114--126.
\end{thebibliography}
\end{document}